\documentclass{ws-ijmpb}

\begin{document}

\markboth{Vladimir Stadnichuk, Anna Bodrova, Nikolai Brilliantov}
{Smoluchowski aggregation-fragmentation equations: Fast numerical algorithm for steady-state solution}

%
\catchline{}{}{}{}{}
%

\title{SMOLUCHOWSKI AGGREGATION-FRAGMENTATION EQUATIONS: FAST NUMERICAL ALGORITHM FOR STEADY-STATE SOLUTION}

\author{VLADIMIR STADNICHUK}

\address{Faculty of Physics, M.V.Lomonosov Moscow State University, Moscow, 119991, Russia\\
stadn@polly.phys.msu.ru}

\author{ANNA BODROVA}

\address{Faculty of Physics, M.V.Lomonosov Moscow State University, Moscow, 119991, Russia\\
bodrova@polly.phys.msu.ru}

\author{NIKOLAI BRILLIANTOV}

\address{Department of Mathematics, University of Leicester, Leicester LE1 7RH, United Kingdom\\
nb144@leicester.ac.uk}

\maketitle

\begin{history}
\received{Day Month Year}
\revised{Day Month Year}
\end{history}

\begin{abstract}We propose an efficient and fast numerical algorithm of finding  a stationary solution of large systems of aggregation-fragmentation equations of Smoluchowski type for concentrations of reacting particles. This method is applicable when the stationary concentrations steeply decreases with increasing aggregate size, which is fulfilled for the most important cases. We show that under rather mild restrictions, imposed on the kernel of the Smoluchowki equation, the following numerical procedure may be used: First, a complete solution for a  relatively small number of equations (a "seed system") is generated and then the result is  exploited in a fast  iterative scheme.  In this way the new algorithm allows to obtain a steady-state solution for rather large systems of equations,  by orders of magnitude faster than the standard schemes.
\end{abstract}


\section{Introduction}
Aggregation and fragmentation  are ubiquitous in nature and many industrial processes are based on these physical  phenomena. As a representative example one can mention reversible polymerization in solutions, coagulation in colloidal solutions,  formation of proto-planets in dust clouds in the inter-galaxy space, etc. The size distribution of the particles  in planetary rings, like Saturn rings, is also determined by a subtle balance between aggregation and fragmentation of the rings objects. Aggregation processes  play an important role in living systems: Coagulation of prions takes place in numerous prion-related diseases; moreover, the blood clotting may be mentioned as the most vivid
example from the everyday life \cite{Krapivsky,Brill2009,Leyvraz}.
For space homogeneous systems, addressed in here the above processes are described by Smoluchowski-like equations. These quantify the change of the size of particles due to formation of joint aggregates from colliding entities, or due to fragmentation of the aggregates. While the former process is necessarily binary, the latter one may be either binary, when particles break upon a collision or unary, when a decay of an aggregate into smaller pieces takes place spontaneously. Binary fragmentation occurs due to inter-particle impacts, when part of the kinetic energy of colliding objects (for instance, of ice grains in planetary rings) exceeds some fragmentation threshold. Unary decomposition may occur due to thermal fluctuations, for example, in the case of decomposition of polymers \cite{Prion}, or in Saturn rings due to meteoroid bombardment. In both cases of spontaneous or collisional fragmentation the distribution of debris has some particular form, depending on the nature of the fragmentation process.

In the present study only binary collisions are addressed, that is, we do not consider inter-particle reactions which involve three or more particles. This is justified for the most important applications; it is a rather seldom process, when an aggregation or fragmentation process may occur if more than two particles are involved.

Below we will study the following fragmentation processes: (i) spontaneous decomposition due to thermal fluctuations (unary process) (ii) fragmentation due to inter-particle impact, when part of the kinetic energy of colliding objects (for instance ice grains in planetary rings) transforms into the surface energy of debris. While in the former case fragmentation occurs as a random event, the latter one is possible  if the kinetic energy of particles exceeds some threshold~\cite{Brill2009}. 

We assume that all aggregates in the system are comprised of particles of minimal size, monomers, which do not suffer further splitting~\cite{Krapivsky,Leyvraz}. The physical nature of the monomers may be very different, depending on the nature of the system. For instance, in polymer solution the role of monomers play some functional chemical groups, which can associate into larger molecules. In planetary rings, the role of monomers play ice grains of a minimal size observed in these system; such grains,  however, may be formed from still smaller particles.

Let $n_k(t)$ be  concentration (a number of particles in a unit volume) of aggregates comprised of $k$ monomers, of mass $m_1$ and radius $r_1$, at time $t$. The Smoluchowski equations describe time evolution of the concentrations $n_k(t)$. As a system evolves, larger and larger aggregates emerge, hence, strictly speaking the system of Smoluchowski equations must be infinite. Obviously, a  direct application of standard numerical schemes to solve such infinite systems is not possible, so we need to develop a more advanced approach. In practice, only a finite number of equations may be numerically solved, therefore a problem arises, how to mimic a behavior of an infinite system with a use of a finite subset; moreover, the basic physical principles, like conservation of mass should not be violated. Next problem is the large number of equations, needed for an adequate modeling, even if a finite system is used. Indeed, the number of different species (that is, aggregates of different sizes) may be very large in some systems; it is equal, for instance, to $10^9-10^{12}$ for planetary rings, where the largest aggregates may contain up to $10^9-10^{12}$ primary particles \footnote{Indeed, while the primary particles have the radius of $\sim 1\,cm$, the maximal aggregate size could be about 5-10 meters \cite{zebker,Cuzzi,Esposito}.} One of the possible solutions is the application of the coarse-graining,  where concentrations of aggregates with close number of monomers are grouped together; the larger is the size of an aggregate, the larger is the respective group~\cite{Malyshkin:2001}. Unfortunately, this approach is not always possible, especially in the case when a sharp cross-over from one functional form of size distribution to another one takes place. As we will show below such cross-over is indeed observed for some of  the studied systems. There are other methods to handle large systems of Smoluchowski equations, e.g.~\cite{Colm:2012,TT-cross,Convolution,VMK}, these methods, are focused, however,  on the time-dependent solutions and do not take precautions against violation of the mass conservation~\cite{TT-cross,Convolution}. The latter is of a primary importance for systems considered here. In the present study we analyze the stationary solutions of large systems of Smoluchowski equations and develop an efficient numerical algorithm. It provides a fast solution for the particle size distribution, even if the concentrations of different species  differ by 10-12 orders of magnitude.  The rest of the paper is organized as follows. In the next Sec.~II we consider the  aggregation-fragmentation equations of the Smoluchowski's type. In the Sec.~III we discuss the most widely used models for the rate coefficients of the Smoluchowski equation and provide the existent exact solution for some particular model. In Sec.~IV we discuss in detail the new numerical approach and present simulation results obtained with the new method. Here we compare the efficiency of our  method with the traditional one. In the last Sec.~V we summarize our findings.

\section{Aggregation-fragmentation models}

To introduce the notations we start with the standard Smoluchowski aggregation equations. The aggregation process may be symbolically written  as
$$
[i] +[j]  \longrightarrow  [i+j] \,.
$$
In what follows we will consider space uniform systems, that is, the local concentrations  $n_i({\bf r},t)$ are assumed  to be uniform trough the system, $n_i({\bf r},t)=n_i(t)$. We also introduce the kinetic coefficients $C_{ij}$, which give the production rate of aggregates of size $i+j$ from particles of size $i$ and $j$. $C_{ij}$ quantifies  the number of such aggregates that appear in a unit volume during a unit time. In these notations the kinetic equations read:
\begin{equation}
\label{eq:1} \frac{dn_k}{dt} = \frac{1}{2} \sum_{i+j=k}C_{ij}n_in_j - \sum_{i=1}^{\infty}C_{ik}n_in_k  \, .
\end{equation}
The equations (\ref{eq:1}) with  $k=1, 2, \ldots$ are the standard Smoluchowski equations. The first term in the r.h.s. of  the above equation describes the rate at which aggregates of size $k$ are formed from particles  of size $i$ and $j$. The summation extends over all $i$ and $j$ with $i+j=k$, and the factor $\frac12$ prevents double counting. The second term gives the rate at which the particles of size $k$ disappear due to aggregation with other particles of any size $i$.

Fragmentation is the opposite process when an aggregate splits into smaller pieces. It may happen
spontaneously, due to thermal fluctuations, when, for instance a particle of size $k$ breaks into $l$
fragments of size $i_1, \, i_2, \, \ldots i_l$,
$$
[k ] \longrightarrow [i_1] +[i_2] + \ldots [i_l] \,
$$
where $i_1 +i_2+ \ldots i_l =k$, due to the mass conservation.

The fragmentation is also possible due to collisions of aggregates. In this case the kinetic energy of their relative motion transforms
into kinetic energy and surface energy of debris. This process may be symbolically written as
$$
[k]+[j] \longrightarrow [i_1] +[i_2] + \ldots [i_l] \,
$$
where again $i_1 +i_2+ \ldots i_l =k+j$ due to the mass conservation. Let us consider a simplified collision model, assuming that particles break completely in a disruptive collision into monomers of mass $m_1$. The kinetic equations, describing all the above processes of aggregation as well as spontaneous and collision fragmentation read,
\begin{eqnarray}
\label{system}
\frac{dn_k}{dt} &=& \frac12 \sum_{i+j=k}  C_{ij} n_in_j - \sum_{i\geq 1} \left(C_{ik}+A_{ik}\right) n_in_k-A_k n_k\\
\frac{d n_1}{dt} &=&   - n_1\sum_{j\geq 1}C_{1j}n_j + n_1\sum_{j\geq 2}jA_{1j}n_j+\frac{1}{2} \sum_{i,j\geq 2}A_{ij}(i+j) n_in_j  + \sum_{j\geq 2}jA_j n_j \,. \nonumber
\end{eqnarray}
Here $A_k$ is the rate of spontaneous fragmentation and $A_{ij}$ - the rate of collisional fragmentation. Note, that similar equations have been studied in the context of of rain drop formation \cite{Srivastava1982}.

The dependence of aggregation and fragmentation rate coefficients on the particles size is determined by the nature of the process. For example, the rate coefficients for the  diffusion-limited aggregation read~\cite{Leyvraz}:
\begin{equation}
\label{eq:diff}
C_{ij} =C_0(i^{1/3} +j^{1/3}) (i^{-1/3} +j^{-1/3})
\end{equation}
where the constant $C_0$ depends on the characteristic size of agglomerates and their characteristic mobility. In the case of ballistic aggregation such dependence has the form~\cite{Brill2009,Palaniswaamy2006}:
\begin{equation}
\label{eq:ballist}
C_{ij} = C_0^{\prime} \left(i^{1/3} + j^{1/3} \right)^{2} \left(i^{-1} + j^{-1} \right)^{1/2}\,,
\end{equation}
where again, $C_0^{\prime} $ is determined by the characteristic size and velocity of the particles. Noteworthy, that the rate coefficients (\ref{eq:diff}) and (\ref{eq:ballist}) are homogeneous functions of masses of  colliding particles $m_i \sim i$ and $m_j \sim j$. Sometimes simplified rate kernels are considered \cite{Leyvraz}:
\begin{equation}
\label{eq:ballist1}
C_{ij} = C_0 \left(i j \right)^{\mu} \, .
\end{equation}

It may be shown that for some important applications the  collision fragmentation coefficients $A_{ij}$ are proportional to the aggregation coefficients,  e.g. ~\cite{PNAS}
\begin{equation}
\label{eq:coll_fragm}
A_{ij}= \lambda C_{ij},
\end{equation}
where $\lambda$ is a constant that characterizes the ratio of the disruptive and aggregative collisions. Usually, it is rather small, $\lambda \ll 1$. For the case of spontaneous fragmentation the following model may be applied,
\begin{equation}
\label{eq:spont_fragm}
A_k =\nu k^{\theta} ,
\end{equation}
where the exponent $\theta$ depends on the dimension of aggregates (which may be, generally fractal) and the average coordination number of the primary particles packing.

In what follows we describe the fast numerical algorithm of finding the steady-state solution of the large sets of Smoluchowski-type equations (\ref{system}). It is also important to check the accuracy of the novel approach. Therefore we give here, without derivation, the exact steady-state solution to these equations for the particular case of constant rate coefficients, $C_{ij}=C_0$, $A_{ij}=\lambda C_0$ with $\lambda ={\rm const}$ and $A_k=\nu$, that is, for $\mu = \theta =0$\cite{Juergen,Bodrova}:
\begin{eqnarray}
\label{eq:n_k}
n_k&=&\frac{1}{\sqrt{4\pi}}\frac{\Gamma\left(k-\frac12\right)}{\Gamma\left(k+1\right)}\left(\frac{2n_1}{(1+\lambda)M_0+\nu}\right)^k \left[\left(1+\lambda\right)M_0+\nu\right]\\
n_1 &=&\frac{M\left(\lambda M_0+\nu\right)}{\nu + (1+\lambda)M_0} \nonumber \\
 M_0&=&\frac{\lambda M - \nu + \left(\left(\nu-\lambda M\right)^{2}+2\nu M\left(1+2\lambda\right)\right)^{1/2}}{1+2\lambda} . \nonumber
\end{eqnarray}
Here $\Gamma(x)$ is the gamma function, $M=\sum_{i=1}^{\infty} k n_k$ is the total mass and $M_0=\sum_{i=1}^{\infty} n_k$ -- the total number of the aggregates; without the loss of generality we take $C_0=A_0=1$.  The derivation details for  the above solution will be given in the forthcoming publication\cite{Bodrova}.

\section{Fast numerical algorithm}
\subsection{Description of the algorithm}
Solving the system of Smoluchovski equation one encounters a number of technical problems. The first problem in the numerical analysis of the system of infinite number of rate equations (\ref{system}) is the conservation of mass of the constituents. Indeed, in any real simulation one can handle only a \emph{finite} number of equations say $N$, which describe evolution of particles of size $1,\, 2,\, \ldots N$ (a particle of size $k$ has mass $m_k=m_1 k$). These equations have
both aggregation and fragmentation terms. In particular, they have a term which describes aggregation of
particles of size $i <N$ and $j <N$, resulting in an aggregate of size $i+j >N$. Since the system of $N$
equations does not account for particles larger that $N$, such processes would effectively lead to the leak
of particles' mass and violation of the mass conservation. To preserve the mass conservation we consider the model, where all collisions with $i+j  >N$ are disruptive, that is, such collisions produce $i+j$ monomers. We have checked that this additional assumption does not lead to any noticeable distortion of the numerical solution of the rate
equations $n_k$,  provided $N$ is large enough and $k$ is smaller than some fraction of $N$.

The most significant problem is to efficiently handle a large number of equations, say up to $\sim 10^{10}$. In principle, this may be done by a coarse-graining, that is, by grouping all concentration  $n_k$ -- $n_{k+l}$ into
coarse-grained variables $\tilde{n}_K$ with increasing $l$ as $k$ grows~\cite{Malyshkin:2001}. In the case of interest, however, we have a drastic variation of the functional dependence of $n_k(k)$, which changes from a power-law to the
exponential decay. This hinders an effective application of the coarse-graining and we need to keep
explicitly all individual concentrations. Hence we have to work with a large number of equations, which is computationally costly. To speed up the computations we  have developed the following recursive procedure for the solution of system of equations (\ref{system}). Taking into account, that we search for a stationary solution, $dn_{k+1}/dt=0$, we obtain  for the number density $n_{k+1}$:
\begin{equation}
\label{eq:begin} \frac{1}{2(1+\lambda)}\sum_{i+j=k+1}C_{ij}n_{i}n_{j}
-\sum_{i=1}^{N}C_{i\,k+1}n_{i}n_{k+1}-\frac{A_k}{1+\lambda}n_k=0\,.
\end{equation}
The first sum in Eq.~(\ref{eq:begin}) contains only $n_i$ with $i\leq k$, while we write the second sum as
\begin{eqnarray}
\sum_{i=1}^{N}C_{i\,k+1}n_{i}n_{k+1} &=&n_{k+1}\sum_{i=1}^{k}C_{i\,k+1}n_{i}+C_{k+1\,k+1}n_{k+1}^2+n_{k+1}\sum_{i=k+1}^{N}C_{i\,k+1}n_{i} \,. \label{eq:javn}
\end{eqnarray}
Now we use the properties of the kinetic kernel $C_{ij}$ and the steady-state distribution $n_k=n_k(m_k)$,
which we assume to be a decreasing function of $k$.  Namely, we assume that the coefficients $C_{ij}$ remain constant or increase with $i$ and $j$ at a smaller rate than the rate at which $n_k$ decreases with $k$. That is, we assume that
for $k \gg 1$ the following condition holds true:
\begin{equation}\label{eq:cond}
\sum_{i=1}^{k}C_{i\,k+1}n_{i}\gg\sum_{i=k+1}^{N}C_{i\,k+1}n_{i} \,.
\end{equation}
This allows to neglect the last sum in Eq.~(\ref{eq:javn}) and obtain the quadratic equation for $n_{k+1}$:
\begin{equation}
\label{eq:sqeq} C_{k+1\,k+1}n_{k+1}^{2} +n_{k+1}\left(\frac{A_{k+1}}{1+\lambda}+\sum_{i=1}^{k}C_{i\,k+1}n_{i}\right)  - \sum_{i+j=k+1}
\frac{C_{ij}n_{i}n_{j}}{2(1+\lambda)}=0 \,.\nonumber
\end{equation}
Solving the above  equation and choosing the positive root, we arrive at the recurrent relation for the concentrations $n_k$:
\begin{eqnarray}
\label{eq:req} &&n_{k+1} = \frac{ \sqrt{D}- \sum\limits_{i=1}^{k}C_{i\,k+1}n_{i} - \frac{A_{k+1}}{1+\lambda}}{2C_{k+1\,k+1}} \\
&&D= \frac{2C_{k+1\,k+1}}{(1+\lambda)}\sum\limits_{i+j=k+1}C_{ij}n_{i}n_{j}+
\left(\sum\limits_{i=1}^{k}C_{i\,k+1}n_{i}+\frac{A_{k+1}}{1+\lambda}\right)^2\,. \nonumber
\end{eqnarray}

\begin{figure}\centerline{\includegraphics[width=0.65\textwidth]{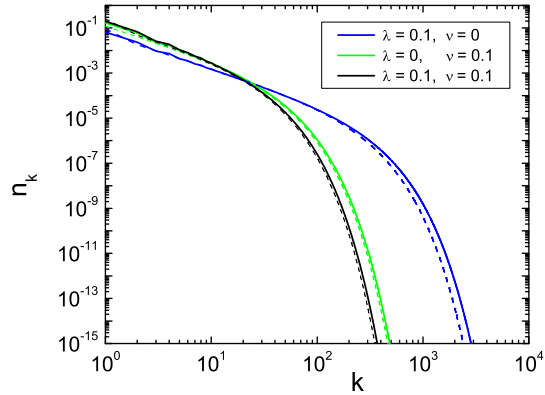}}
\caption{ Equilibrium size distribution of particles $n_k$ for $C_{ij}=1$, $A_{ij}=\lambda$, $A_k=\nu$. The total number of equation is $N=16000$ and the size of the "seed system" is  $K=1500$ (see the text for the explanations). A good agreement between the numerical (solid lines) and exact solution, Eq.~(\ref{eq:n_k}) (dashed lines) is observed. A slight deviation from the exact solution for large $k$ may be attributed to the computational errors, when very small numbers are handled.} \label{Gconst}
\end{figure}
Using the recurrent relation (\ref{eq:req}) one can significantly accelerate computations. This may be done as follows:~First, we solve explicitly the system of $K$ rate equations (\ref{system}) for $K \ll N$, choosing the value of $K$ to fulfill the condition (\ref{eq:cond}). Next,  the concentrations $n_i$ with $K < i \leq N$ may be straightforwardly obtained from the recurrent relation (\ref{eq:req}). In other words, our new algorithm uses the following trick: First we "exactly" solve a relatively small "seed system" and than apply the recurrence to obtain the solution for the whole system.

Solving numerically the rate equations with different kernels directly, and with the use of our recurrent method, we proved the efficiency and accuracy of the above accelerating approach. The computation accuracy may be checked by a  direct comparison of the numerical and exact solutions.  The results of numerical solution of system of equations (\ref{system}) is given in Fig.~\ref{Gconst}. A good agreement with the theory is observed.

\subsection{Analysis of the algorithm efficiency}
The efficiency of our new algorithm can be evaluated in the following way. The direct solution of the system of equations requires  $\sim N^2$ operations at each time step. Let $T$ be the number of time steps, needed to arrive at the steady state. Since the relaxation time depends only on the physical parameters of the system, $T$ should not be sensitive to the number of the equations $N$. Hence, it is reasonable to approximate it by a constant, $T(N) ={\rm const} \gg 1$. Then the number of operations $t(N)$, required to find a steady-state solution of a system of $N$ equations reads, $t(N) = T N^2$. The number of operations $t(N)$ refers to the standard computation scheme.

Consider now the number of operations needed in our new approach for a system of $N$ equation. Let we start with the "seed system" of $K$ equations. To find a  steady-state solution for the first $K$ concentrations we need $TK^2$ operations. Then to obtain the steady state solution of the $(K+1)$-st concentration $n_{K+1}$, using the fist $K$ concentrations $n_1, \, n_2, \, \ldots, n_K$, one needs $K+1$ operation, see Eq.~(\ref{eq:req}). For the next $(K+2)$-th concentration, $(K+2)$ operations are required, etc. Hence, to find the rest $(N-K)$ steady-state concentrations one has to perform
$$
(K+1)+(K+2)  \ldots + N = \frac12(N-K)(N+K-1) \simeq \frac12 (N^2-k^2)
$$
operations, so that $TK^2 + (N^2-k^2)/2$ operations in total are needed to obtain the solution within the new approach. Then the ratio of computation times, required for the direct (standard) and new algorithm reads:
\begin{equation}
\frac{t_{\rm rec}}{t_{\rm stan}}\simeq
\frac{TK^2+(N^2-K^2)/2}{TN^2}=\frac{K^2}{N^2}+\frac{1}{2T} -\frac{K^2}{2N^2 T} \sim\frac{K^2}{N^2}\, ,
\end{equation}
where we use the fact that $N \gg K$ and that the number of time steps, $T \gg 1$, which refers to the relaxation time, is always in practice much larger  that the ratio $N^2/K^2$.

\begin{table}[ph]
\tbl{Comparison of the CPU time of the direct (standard) and new algorithms for the solution of the Smoluchowski equations for the case of collisional fragmentation for different number of equations. $C_{ij} =1$,  $A_{ij} =\lambda$ and  $A_k=0$}
{\begin{tabular}{|c|c|c|c|c|c|}\hline
$\lambda$&$K$&$N$&$t_{\rm rec}$(a.e.)&$t_{\rm stan }$(a.e.)&$t_{\rm stan}/t_{\rm rec}$\\ \hline
$0.1$&$10^3$&$5\cdot 10^3$&37&1787&48\\ \hline
$0.05$&$10^3$&$10^4$&38&10800&284\\ \hline
$0.01$&$1.5\cdot 10^3$&$10^4$&80&10090&126\\ \hline
\end{tabular}}
\label{tab1}
\end{table}

In Table~\ref{tab1} we present the comparison of the efficiency (in terms of the CPU time) of our new algorithm and of the standard one; as it may be seen from the table, the new algorithm is indeed much faster and efficient.  Note that the significant acceleration in the systems solutions is obtained without any loss in the computation  accuracy.

\section{Conclusion}
We perform a numerical analysis of Smoluchowski-like aggregation-fragmentation equations to find a steady-state size distribution of particles. We address very large systems of equations as  required in  some important applications. We develop a new  numerical algorithm which yields a significant acceleration of simulations without the loss of its accuracy. We compare our method to the standard one and confirm its impressive superiority in speed.  The developed algorithm may be exploited not only for the Smoluchowski-like aggregation-fragmentation equations but for other set of equations which have the structure similar to that addressed in our study.

\section*{Acknowledgements} AB gratefully acknowledges the use of the facilities of the Chebyshev supercomputer of the Moscow State University.

\end{document}